\documentclass[prl,twocolumn,aps,superscriptaddress]{revtex4}

\usepackage{graphicx}% Include figure files
\usepackage{dcolumn}% Align table columns on decimal point
\usepackage{bm}% bold math
\usepackage{amssymb}

\begin{document}

%\preprint{Preprint}

\title{Kitaev-Heisenberg model in a magnetic field: order-by-disorder and commensurate-incommensurate transitions}

\author{Gia-Wei Chern}
\affiliation{Department of Physics, University of Virginia, Charlottesville, VA 22904, USA}

\author{Yuriy Sizyuk}
\affiliation{School of Physics and Astronomy, University of Minnesota, Minneapolis, MN 55455, USA}

\author{Craig Price}
\affiliation{Department of Physics, The Pennsylvania State University, University Park, PA 16802, USA}

\author{Natalia B. Perkins}
\affiliation{School of Physics and Astronomy, University of Minnesota, Minneapolis, MN 55455, USA}

\date{\today}

\begin{abstract}
We present a theoretical study of field-induced magnetic phases in the honeycomb Kitaev-Heisenberg model, which is believed to describe the essential physics of Mott insulators with strong spin-orbit coupling such as $A_2$IrO$_3$ and $\alpha$-RuCl$_3$. We obtain a finite temperature phase diagram based on extensive Monte Carlo simulations and analytical calculations. We show that, while Zeeman coupling favors a symmetric non-coplanar magnetic order, thermal fluctuations enhances the stability of a collinear zigzag phase that breaks the rotational symmetry of the lattice. Our large-scale simulations also uncover intriguing commensurate-incommensurate transitions and multiple-$\mathbf Q$ incommensurate phases at high field. Experimental implications are also discussed. 
\end{abstract}

\maketitle

Recently, a great interest has emerged in the  study of magnetic properties of 4$d$ and 5$d$ transition metal  systems such as iridates and ruthenates~\cite{rau16}.
%This interest is partly driven by the realization that the Kitaev model might be realized in these compounds with anisotropic exchange interactions~\cite{jackeli09}.
%The Kitaev model is an exactly solvable quantum spin-1/2 system whose ground state is a novel quantum spin liquid with fractionalized excitations~\cite{kitaev06}. 
In comparison with 3$d$ compounds, the iridates and ruthenates have weaker Coulomb correlations but a much stronger relativistic spin-orbit coupling. The latter entangles the spin and orbital degrees of freedom into an effective total angular moment, which in the case of Ru$^{3+}$ and Ir$^{4+}$ is $J_{\rm eff}=1/2$. Notably, the orbital character of the pseudospin gives rise to highly anisotropic and spatially dependent exchange interactions.
Significant experimental effort has been devoted to study systems in which these magnetic ions occupy sites with three-fold coordination in a structure with edge-sharing octahedra~\cite{singh10,singh12,liu11,ye12,choi12,gretarsson13,williams16,modic14,biffin14-1,biffin14-2,tomo15,plumb14,sears15, majumber15,johnson15,banerjee15}. 
%Notable examples include the well-studied honeycomb iridates $A_2$IrO$_3$ ($A$ = Na, Li)~\cite{singh10,singh12,liu11,ye12,choi12,gretarsson13} and the ruthenium chloride RuCl$_3$~\cite{plumb14,sears15,majumber15,johnson15,banerjee15} in two dimensions. There are also three-dimensional systems such as the hyperhoneycomb or multilayer-honeycomb iridates~\cite{modic14,biffin14-1,biffin14-2,tomo15}.

The motivation behind this flurry of experimental activity is the possibility of realizing the  Kitaev quantum spin liquid~\cite{jackeli09}, because this lattice geometry promotes the dominance of the Kitaev interactions between their magnetic moments. The Kitaev model is an exactly solvable quantum spin-1/2 system whose ground state is a novel quantum spin liquid with fractionalized excitations~\cite{kitaev06}.
However, it was soon realized that  at sufficiently low temperatures all  these compounds order magnetically  rather than  exhibiting  spin-liquid  behavior. These findings suggested the importance of other  subdominant interactions between magnetic moments in these spin-orbital coupled Mott insulators~\cite{jackeli10,jackeli13,sizyuk14,katukuri14,rau14,yamaji14,kee15,roser16,kee2016,chaloupka15,ioannis15,sizyuk16,chaloupka16}.

A particularly important interaction in addition to the Kitaev coupling is the isotropic nearest-neighbor (NN) Heisenberg exchange due to direct overlap of the $d$ orbitals. The frustrated nature of spin interactions in this so-called Kitaev-Heisenberg (KH) model manifests itself in the many competing magnetic orders as well as two quantum spin liquids in the phase diagram~\cite{jackeli13}. Frustration also means that the system is sensitive to the perturbation of a magnetic field~\cite{vojta16}. Indeed, novel magnetic phases such as fractional magnetization plateaus or skyrmion crystals can be stabilized by a magnetic field in several highly frustrated magnets.

%The frustrated nature of the spin interactions manifests itself in the fact that already the combination of the Kitaev interaction and the

In this paper, we discuss  field-induced phenomena in the honeycomb KH model based on a complete temperature-field phase diagram obtained from our extensive Monte Carlo simulations. We focus on the zigzag phase which is relevant for Na$_2$IrO$_3$ and $\alpha$-RuCl$_3$, and the $[111]$ magnetic field direction such that the discrete rotational symmetry of the lattice is preserved. In the zero temperature limit, which has been studied in Ref.~\onlinecite{vojta16}, our results are consistent with theirs in the case of commensurate phases. Moreover, we have uncovered intriguing discontinuous commensurate-incommensurate transitions and novel triple-$\mathbf Q$ incommensurate zigzag states at high magnetic field.

We consider the  KH model subject to a magnetic field 
\begin{eqnarray}
	\mathcal{H} = J \sum_{\langle ij \rangle} \mathbf S_i \cdot \mathbf S_j  + 2 K \sum_{\langle ij \rangle_\gamma} S^\gamma_i S^\gamma_j - \mathbf H \cdot \sum_i \mathbf S_i.
\end{eqnarray}
Here $\gamma = x$, $y$, and $z$ denote the three distinct  NN bonds of a honeycomb lattice. The spin quantization axes are taken along the cubic axes of the IrO$_6$ octahedra. The first $J$ term is the isotropic Heisenberg exchange, while the second Kitaev term describes the bond-dependent Ising coupling between spin components.
%We focus on the case in which the magnetic field is applied along the $[111]$ direction.
%, such that symmetries related to the $C_3$ rotation are preserved. 

\begin{figure} [b]
\centering
\includegraphics[width=1.01\columnwidth]{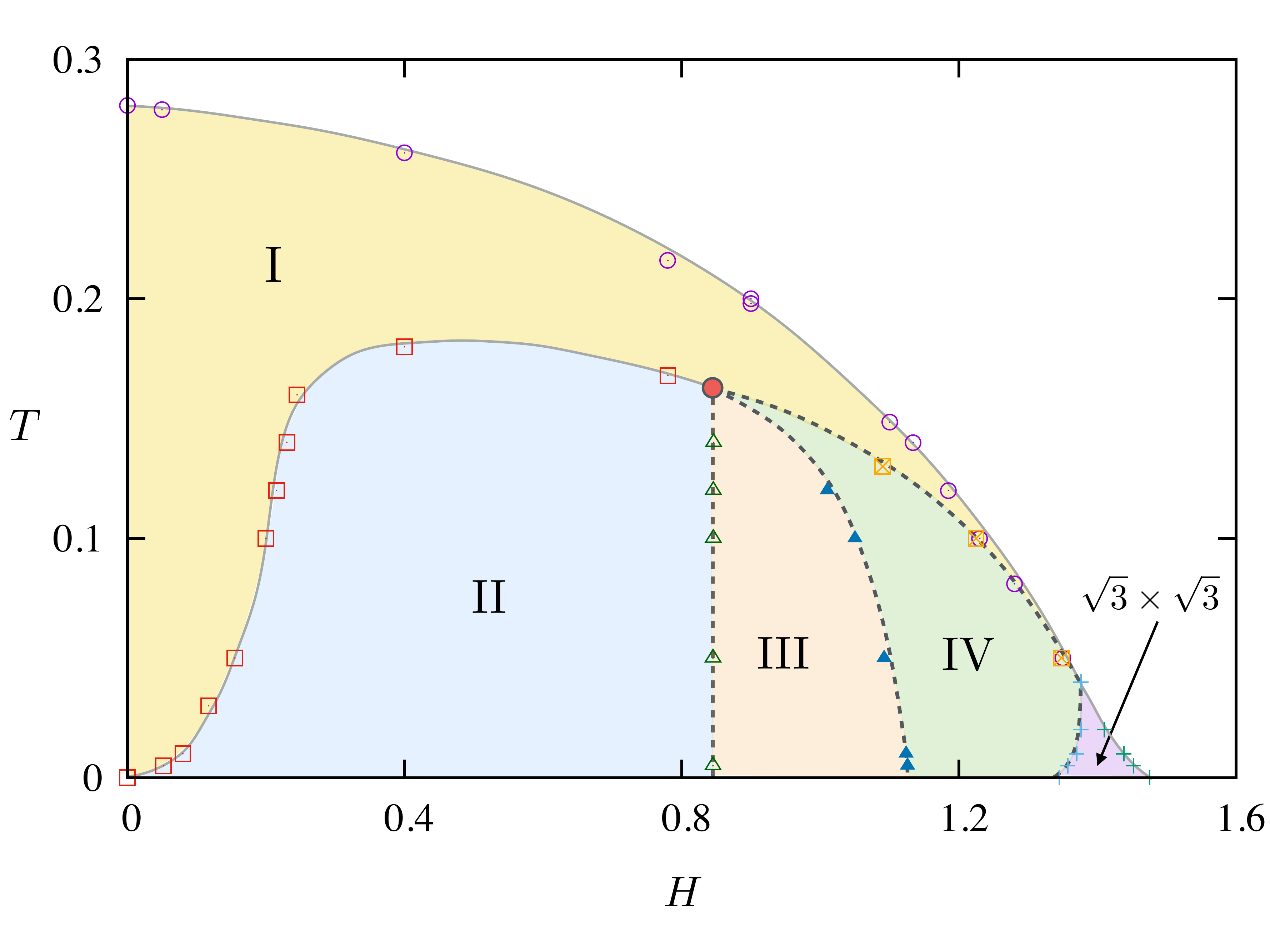}
\caption{\label{fig:HT_diagram}
The field-temperature phase diagram of the KH model with parameter $\varphi = 0.7\pi$. Dashed and solid lines denote first and second-order phase transitions, respectively. There are five ordered phases at low temperatures. Other than the $\sqrt{3}\times \sqrt{3}$ order at high field, the phase diagram is dominated by four distinct zigzag phases: single-$\mathbf Q$ canted zigzag (I), triple-$\mathbf Q$ commensurate zigzag (II), triple-$\mathbf Q$ partial incommensurate zigzag (III), and fully incommensurate 3$\mathbf Q$ zigzag (IV). The corresponding structure factors and spin snapshots are shown in Fig.~\ref{fig:phases}.  $T$ and $H$ are measured in units of overall exchange energy scale $A$.
}
\end{figure}

Already at zero field, the KH model exhibits several interesting phases depending on the relative strength of the two competing terms. A convenient parameterization is to write $J = A \cos\varphi$ and $K = \sin\varphi$, where $A > 0$ is the overall energy scale of exchange interaction. In addition to the conventional ferromagnetic and N\'eel orders, the classical phase diagram includes two collinear antiferromagnetic (AF) states with spontaneously broken $C_3$ symmetry, called the zigzag and stripy AF orders. 
Remarkably, all magnetic phases survive quantum fluctuations and remain stable in the limit of $S = 1/2$, except for two small regions of $\varphi$ close to $\pi/2$ and $3\pi/2$ where quantum spin liquids emerge as the ground states.

\begin{figure*}
\centering
\includegraphics[width=2\columnwidth]{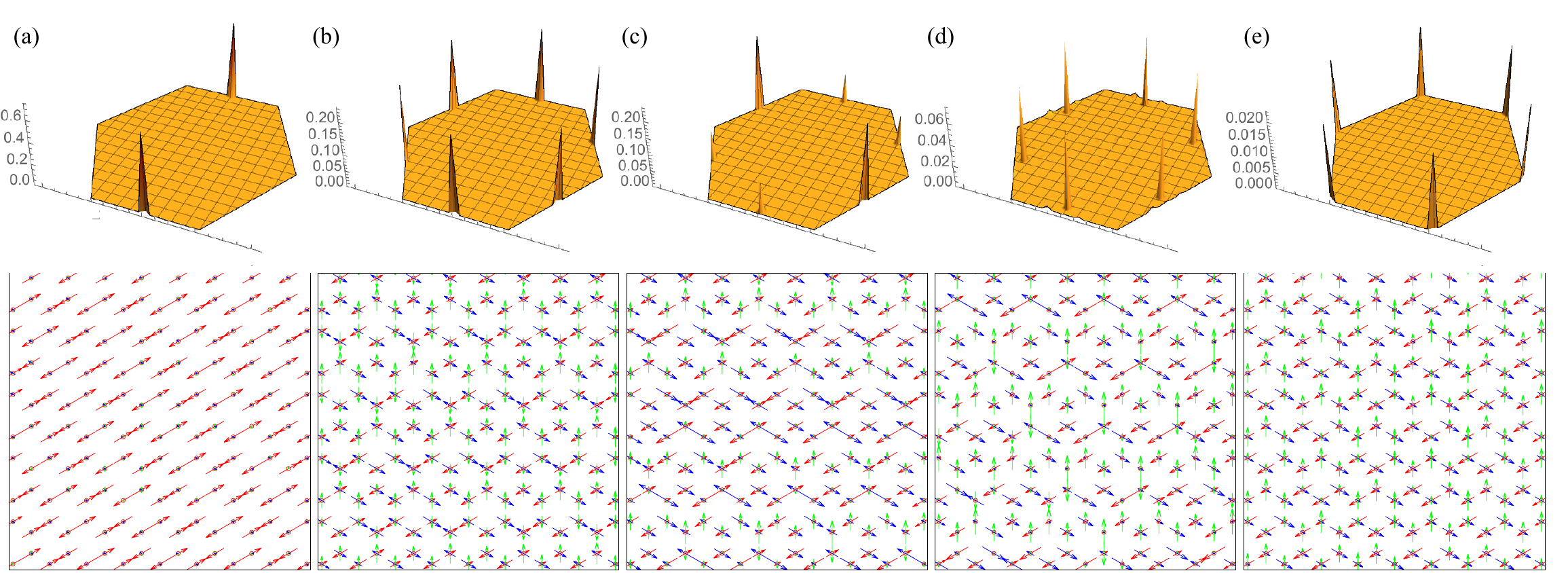}
\caption{\label{fig:phases}
Magnetic phases in the honeycomb KH model. Top row shows the spin structure factor obtained from simulations at $T = 0.005$; Corresponding snapshots of the spin configurations are shown in the bottom row. The three spin components are shown here with red, green, and blue colors. The five phases shown here are (a) single-$\mathbf Q$ collinear zigzag order ($H = 0.016$), (b) commensurate triple-$\mathbf Q$ non-coplanar zigzag ($H = 0.48$), (c) coexistence of commensurate and incommensurate triple-$\mathbf Q$ zigzag phase  ($H = 1.02$), (d) incommensurate triple-$\mathbf Q$ zigzag phase ($H = 1.34$), and (e) $\sqrt{3} \times \sqrt{3}$ order ($H = 1.4$). $T$ and $H$ are measured in units of $A$.
}
\end{figure*}

The zigzag phase, which is our primary interest in this work, occupies almost a quarter of the phase space~($0.501\,\pi \lesssim \varphi \lesssim 0.9\,\pi$) of the KH model at $T = H = 0$~\cite{jackeli13}.
Here we focus on the KH model with parameter $\varphi = 0.7\pi$ and employ Monte Carlo simulations to study the $H$-$T$ phase diagram. Our extensive simulations result in an unexpectedly rich phase diagram shown in Fig.~\ref{fig:HT_diagram}, which is dominated by four distinct zigzag phases labeled by I, II, III, and IV. In addition, a non-collinear $\sqrt{3}\times \sqrt{3}$ order is stable in a  magnetic  field just below the saturation  and low temperature regime. The representative snapshots and the corresponding spin structure factors of these five ordered phases are shown in Fig.~\ref{fig:phases}. In the following, we discuss the properties of these phases and their numerical characterizations. 

We begin with the single-$\mathbf Q$ zigzag order (phase I), which is the low-$T$ phase of the KH model at $H = 0$.
This ordered state is characterized by collinear spins forming ferromagnetic zigzag chains, which are anti-collinearly staggered along the direction perpendicular to the chains; see Fig.~\ref{fig:phases}(a). Importantly, the direction of collinear spins is locked to orientation of the zigzags.
There are three degenerate zigzag states that are related to each other by symmetry; they correspond to the three staggering wavevectors: $\mathbf Q_{1,2} = (\pm \pi, -\pi/\sqrt{3})$, and $\mathbf Q_3 = (0, 2\pi/\sqrt{3})$, which are the middle $M$ points of the Brillouin zone (BZ) edges.  
The collinear zigzag phase can be characterized by an Ising order parameter $\phi_m$, which is the odd-parity one-dimensional irreducible representation of the little group corresponding to wavevector $\mathbf Q_m$. A general multiple-$\mathbf Q$ zigzag state is then described by a pseudo-vector of three Ising parameters: $\bm \phi = (\phi_1, \phi_2, \phi_3)$.
In terms of the triplet order parameter, the spins in a general zigzag state are expressed as $S^\gamma_i = \pm \phi_m \, S \exp(i \mathbf Q_m \cdot \mathbf r_i)$; where $\pm$ is used for the two sublattices of honeycomb, and the spin component $\gamma = x, y, z$ corresponds to $m = 1, 2, 3$, respectively.  

In the framework of the Ginzburg-Landau theory, the transition into the zigzag phase is described by a free-energy expansion in terms of the pseudo-vector order parameter~$\bm\phi$. Up to quartic order, it reads:
\begin{eqnarray}
	\label{eq:F0}
	\mathcal{F} = r |\bm\phi|^2 + u |\bm\phi|^4 + g\, \phi_1 \phi_2 \phi_3  + v \sum_{m\neq n} \phi_m^2 \phi_n^2.
	%\nonumber \\ + v (\phi_1^2 \phi_2^2 + \phi_2^2 \phi_3^2 + \phi_3^2 \phi_1^2). 
\end{eqnarray}
While this free energy respects the $C_3$ symmetry of the KH model, the first two terms actually preserve a $O(3)$ rotational symmetry of the pseudo-vector $\bm\phi$, indicating an emergent continuous degeneracy of the zigzag states. Indeed, explicit calculation shows that all multiple-$\mathbf Q$ zigzag states satisfying $|\bm\phi| = $ constant are degenerate at the mean-field level~\cite{jackeli10,sizyuk16}. This accidental degeneracy is lifted by the cubic $g$ and quartic $v$ terms of Eq.~(\ref{eq:F0}). In the absence of magnetic field, the cubic term is not allowed by time-reversal symmetry. On the other hand, thermal and quantum fluctuations select the collinear single-$\mathbf Q$ zigzag order~\cite{price12,sela14}. This order-by-disorder phenomenon indicates a repulsive interaction $v \sim v_{0} + v_1 T$ with $v_{0, 1} > 0$; the two terms corresponds to quantum and thermal contributions, respectively.

%Physically, the accidental continuous degeneracy of the zigzag states results from the frustrated exchange interactions of the KH model. It also indicates that the phase might be very sensitive to magnetic field. To this end, we employ large-scale Monte Carlo simulations to investigate the thermodynamic phases induced by the external field along the [111] direction. 

On the other hand, a finite $g$ is allowed when the time-reversal symmetry is explicitly broken by a magnetic field. This cubic interaction term favors a zigzag order with coexisting $\phi_m$, irrespective of the sign of~$g$. Our Monte Carlo simulations indeed find a triple-$\mathbf Q$ zigzag order (phase II) that is favored by the cubic term in a large portion of the phase diagram; see Fig.~\ref{fig:HT_diagram}. The spin configuration of the triple-$\mathbf Q$ zigzag corresponding to a pseudo-vector $\bm\phi \propto (1, 1, 1)$ is shown in Fig.~\ref{fig:phases}(b). The three spin components participate in ordering along different zigzag directions characterized by the three wavevectors~$\mathbf Q_m$, giving rise to a non-coplanar magnetic structure. Our variational calculations based on a quadrupled unit cell, which encompasses general zigzag patterns, also verifies that the triple-$\mathbf Q$ zigzag state is energetically favored by  any finite $H$~\cite{supplementary}.

\begin{figure}
\centering
\includegraphics[width=0.85\columnwidth]{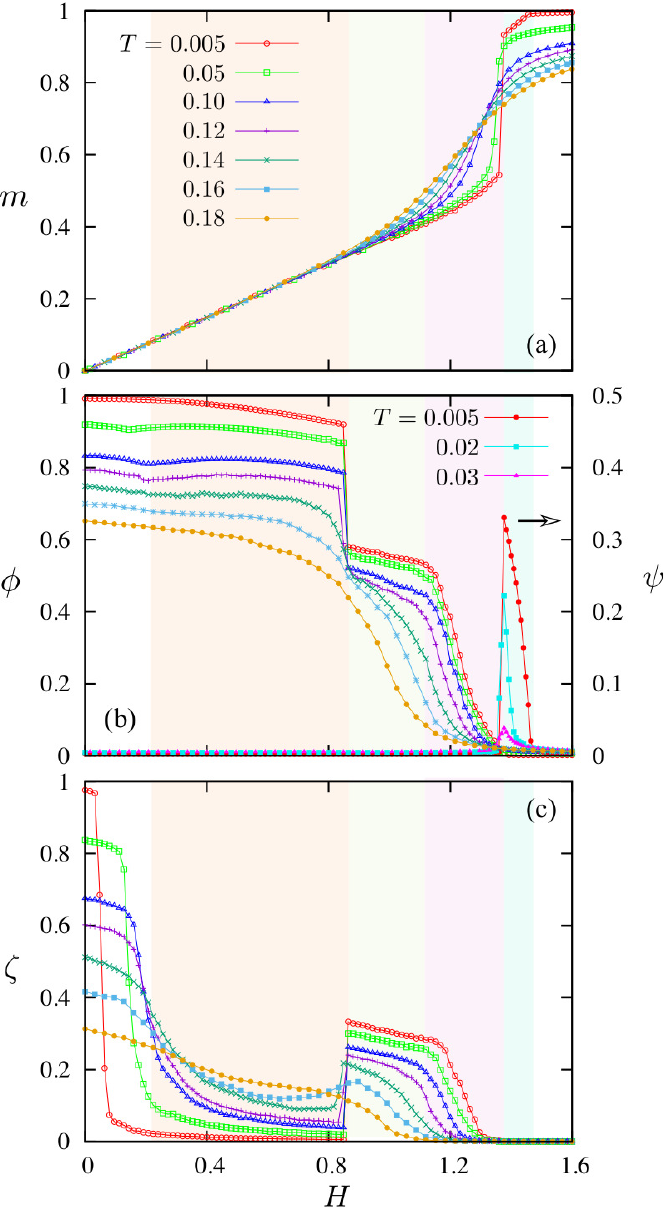}
\caption{\label{fig:h_dep}
Monte Carlo simulations of KH model subject to a magnetic field along the $[111]$ symmetric direction. (a) Magnetization projected onto the field direction as a function of $H$ for varying temperatures. (b) Amplitude of zigzag order parameter $\phi = |\bm \phi|$ (left axis) and the $\sqrt{3} \times \sqrt{3}$ order parameter $\psi$ (right axis) versus field strength. (c) Field dependence of the order parameter $\zeta = |\bm\zeta|$ characterizing the disparity of the three zigzags. Both temperature $T$ and field strength $H$ are in units of the exchange energy scale $A$. The simulations are performed on the KH model with parameter $\varphi = 0.7 \pi$, where single-$\mathbf Q$ collinear zigzag order is the ground state. The number of spins is $N_s = 2 \times 60^2$.
}
\end{figure}

The transition between phases I and II results from the competition between the $v$ and $g$ terms in $\mathcal{F}$, i.e., between the entropic selection and Zeeman energy gain. As the system crosses this phase boundary from the low field side, the broken $C_3$ symmetry of phase I is restored. Interestingly, this phase transition has almost no noticeable effects on the magnetization curve, as shown in Fig.~\ref{fig:h_dep}(a). While a clear jump at high field in the low-$T$ curves indicates a first-order transition into the $\sqrt{3}\times \sqrt{3}$ phase, the magnetization  increases smoothly with $H$ in the small to intermediate field regime. On the other hand, the field dependence of the zigzag order amplitude $|\bm\phi|$, shown in Fig.~\ref{fig:h_dep}(b), exhibits a small kink and a conspicuous drop at intermediate fields, respectively, indicating hidden phase transitions in the seemingly linear magnetization curves. 
%The various field-induced phases along with their structure factors obtained from our simulations are summarized in Fig.~\ref{fig:phases}; below we discuss in detail their structures and characterizations. 

To distinguish the various zigzag orders and particularly to quantify the broken $C_3$ symmetry, we introduce a doublet order parameter $\bm\zeta$ with components:
\begin{eqnarray}
	\zeta_1 = (\phi_1^2 + \phi_2^2 - 2 \phi_3^2)/\sqrt{6}, \quad \zeta_2 = (\phi_1^2 - \phi_2^2)/\sqrt{2},
\end{eqnarray}
which characterizes the disparity of the three zigzag patterns. Physically, a nonzero $\bm\zeta$ corresponds to a spontaneously broken~$C_3$ symmetry. As discussed above, thermal fluctuations at zero field select one of the three collinear zigzag orders, giving rise to a large~$|\bm\zeta|$, while the doublet parameter vanishes in the symmetric triple-$\mathbf Q$ zigzag phase at low temperatures. Indeed, as shown in Fig.~\ref{fig:h_dep}(c), the amplitude of the doublet order parameter decreases with increasing field strength, signaling a transition into a more symmetric zigzag phase.

At intermediate field strength, our Monte Carlo simulations uncover another phase transition hidden in the seemingly smooth magnetization curve. As shown in Fig.~\ref{fig:h_dep}(b) and (c), both order parameters $\bm\phi$ and $\bm\zeta$  exhibit a pronounced discontinuity at $H \sim 0.8$  for temperatures $T \lesssim 0.02$. In particular, the sudden increase of $|\bm \zeta|$ indicates that the $C_3$ symmetry is again broken when crossing this first-order transition from the low-field side. Detailed examinations show that this new zigzag order (phase III in Fig.~\ref{fig:HT_diagram}) is a novel partially incommensurate (IC) phase. Its spin structure factor, shown in Fig.~\ref{fig:phases}(c), exhibits four peaks at IC wavevectors close to the $M$ points, along with two larger peaks remaining at the midpoints of the BZ edges. 

The IC zigzag order can be understood as the corresponding order parameter acquiring a long-wavelength modulation, i.e. $\phi_m (\mathbf r) \sim \cos(\mathbf k_m \cdot \mathbf r + \theta_0)$, where $\theta_0$ is a constant phase, $\mathbf k_m = \varepsilon\, \mathbf Q_m$ is parallel to the corresponding zigzag wavevector and $\varepsilon \ll 1$. The corresponding spin component thus has a spatial dependence: $S^\gamma_i \sim e^{i \mathbf Q_m \cdot \mathbf r_i} \cos(\mathbf k_m \cdot \mathbf r_i + \theta_0)$. In momentum space, since $\mathbf Q_m \equiv -\mathbf Q_m$ up to a reciprocal lattice vector, the single peak at the original commensurate $M$ point splits into two IC peaks at $(1 \pm \epsilon) \mathbf Q_m$.

In phase III, two of the zigzag order parameters, say $\phi_1$ and $\phi_2$, undergo this modulation instability while the third one $\phi_3$ remains commensurate. This asymmetry is responsible for the broken $C_3$ symmetry. In real space, this phase exhibits a stripy superstructure on top of the underlying zigzag pattern. 
As the field is further increased, the remaining commensurate zigzag parameter also undergoes a C-IC transition, giving rise to a fully IC state corresponding to phase IV in Fig.~\ref{fig:HT_diagram}. As shown in Fig.~\ref{fig:phases}(d), the structure factor of this fully IC zigzag exhibits six peaks at momenta that are close to the $M$ points, but inside the BZ. This second C-IC transition is also marked by the decrease of the $\zeta$ parameter, hence partially restoring the $C_3$ symmetry of the system; see Fig.~\ref{fig:h_dep}(c). 

  The  observed C-IC transitions might be partially driven by entropic selection.
 Since thermal fluctuations tend to favor collinear spin configurations,  one of the reasons behind the stabilization of the IC  order can be due to the increase of spin collinearity. Indeed, we  found that
  the IC zigzag state has a larger value of the nematic order than in the  triple-$\mathbf Q$ zigzag phase~\cite{supplementary}. Phenomenologically, these two C-IC transitions result from the softening of the gradient terms of the zigzag order parameters. We can again understand the nature of these two transitions from the Ginzburg-Landau formalism. For convenience, we introduce a triplet of order parameters $\bm\xi = (\xi_1, \xi_2, \xi_3)$ which measure the incommensurability of the corresponding zigzag ordering. More specifically, we define $\xi_m = \hat{\mathbf Q}_m \cdot \nabla \phi_m$. Note that modulations of $\phi_m$ that are perpendicular to $\mathbf Q_m$ are not considered here, since they are not observed in our simulations. Up to the sixth-order, the free-energy of the gradient terms reads
\begin{eqnarray}
	\label{eq:F_grad}
	\mathcal{F}_{\rm grad} = a |\bm\xi|^2 + b |\bm\xi|^4 + c |\bm \xi|^6 +  d \sum_{m\neq n} \xi_m^2 \xi_n^2 + e \, \xi_1^2\, \xi_2^2\, \xi_3^2. \,\,
\end{eqnarray}
Interestingly, the conventional scenario in which the IC phase is caused by the softening of the stiffness constant $a \to 0$ would lead to a continuous phase transition in the Landau theory. Moreover, the quartic interaction term will immediately select a zigzag state with either a single IC zigzag ($d > 0$) or a fully IC  zigzag ($d < 0$). These results are inconsistent with our numerical simulations. Instead, the observed discontinuous C-IC transitions can be attributed to a negative quartic term $b < 0$ while $a$ remains positive throughout the transitions, a scenario similar to the first-order transition close to a tricritical point~\cite{chaikin95}. Here a sixth-order term with $c > 0$ is required for stability of the system. 

The first three terms preserve a pseudo-$O(3)$ rotational symmetry of the modulation parameters $\bm\xi$. Similar to the free-energy in Eq.~(\ref{eq:F0}), this symmetry indicates a continuous degeneracy of IC zigzag orders. The exact  IC order is determined by the interactions among the $\xi_m$ parameters, which are represented by the last two terms in $\mathcal{F}_{\rm grad}$.  A dominant $e>0$, corresponding to a strong repulsion between the modulation parameters, favors the partially IC phase~III in which one of the three $\xi_m$ is zero. On the other hand, a large attractive interaction among the modulations $\xi_m$, represented by a $d < 0$ term, would drive the system into a fully IC  state with restored $C_3$ symmetry. 

At large magnetic field, the IC zigzag phase is connected to a $\sqrt{3}\times \sqrt{3}$ order through another first-order transition, which manifests itself in the huge jump in magnetization at $H \sim 1.35$ at low temperatures. This phase is characterized by a Bragg peak at the $K$ point of the BZ, which also serves as the relevant order parameter. A clear jump of the $\sqrt{3}\times \sqrt{3}$ order parameter $\psi$ can be seen in Fig.~\ref{fig:h_dep}(b). Explicit stability analysis of the fully polarized state indeed shows that the magnetic instability occurs at the $K$ points of the BZ when field is lowered below the saturation field~\cite{supplementary,vojta16}, consistent with our numerical results. 
 
To summarize, we have investigated the finite temperature phase diagram of the KH model subject to a magnetic field. Our extensive Monte Carlo simulations have uncovered several novel zigzag orders and phase transitions. Of particular interest is the existence of two intriguing IC zigzag orderings at intermediate to large field regime. Interestingly, these unusual zigzag states are completely hidden in the magnetization measurement, which shows a smooth growth of magnetic moment with increasing field. 
These intriguing IC zigzags  might be identified in  high-field  $\mu$SR experiments  which provide a powerful means of measuring the internal magnetic field 
distribution caused by the presence of the peculiar field
texture.
%These intriguing IC zigzags might distinguish themselves in high-field $\mu$SR experiments. 
%The C-IC transition could be observed experimentally using  inelastic neutron scattering techniques. 
Finally, although zigzag phases have been detected in Na$_2$IrO$_3$ and $\alpha$-RuCl$_3$, the spin Hamiltonian of both compounds involve further neighbor isotropic and anisotropic interactions. On the other hand, given the frustrated nature of spin interactions in such spin-orbit Mott insulators, we expect similar field-induced phases to occur in real materials, which is left for future studies.

{\em Acknowledgement}. The authors thank C. D. Batista, G. Jackeli, I. Rousochatzakis, and P. W\"{o}lfle for insightful discussions.
 %We also thank C. Price for his help at the early stage of this work. 
N.P. and Y.S. acknowledge the support from NSF Grant DMR-1511768. G.-W. C. and N. B acknowledge the hospitality of Aspen Center for Physics where this work was initiated.

\newpage 

\begin{center}
{\bf Supplementary Material}
\end{center}

\bigskip

In this Supplementing material we provide auxiliary information, some technical details and derivations.
Specifically, Sec.~I  gives details of the classical instability analysis at the saturation field. Sec.~II presents a variational calculation for the classical ground states of the KH Hamiltonian.  In Sec.~III, we characterize the various zigzag phases using the nematic order parameter.  Finally, in Sec. IV  we discuss the nature of the field induced  phase transitions  based on  the annealing and heating simulations.

\begin{center}
{\bf I. Classical instability analysis}
\end{center}

Here we analyze the magnon instability of Kitaev-Heisenberg (KH) model at high magnetic field. Specifically, a linear stability analysis is employed to find the most unstable normal mode of the KH Hamiltonian in a magnetic field. The Hamiltonian of KH model on a honeycomb lattice reads:
\begin{eqnarray}
	\mathcal{H} = J \sum_{\langle ij \rangle} \mathbf S_i \cdot \mathbf S_j  + 2 K \sum_{\langle ij \rangle_\gamma} S^\gamma_i S^\gamma_j - \mathbf H \cdot \sum_i \mathbf S_i.
\end{eqnarray}
We focus on the case in which the field $\mathbf H$ is along the symmetric $[111]$ direction. In the large field limit, all spins are polarized: $\mathbf S_i = S\,\hat{\mathbf n}$, where $\hat{\mathbf n}=\hat{\mathbf e}_{[111]}$ is a unit vector pointing along the [111] direction. For convenience, we will set $S = 1$ in the following discussion. We next introduce two unit vectors $\hat{\mathbf e}_a = (\hat{\mathbf e}_x + \hat{\mathbf e}_y - 2 \hat{\mathbf e}_z)/\sqrt{6}$ and $\hat{\mathbf e}_b = (\hat{\mathbf e}_y - \hat{\mathbf e}_x)/\sqrt{2}$, where $\hat{\mathbf e}_{x, y, z}$ are unit vectors pointing along the three cubic axes. The three vectors $\hat{\mathbf e}_a$, $\hat{\mathbf e}_b$ and $\hat{\mathbf n}$ form an orthonomal basis.

\begin{figure}
%\centering
\includegraphics[width=0.95\columnwidth]{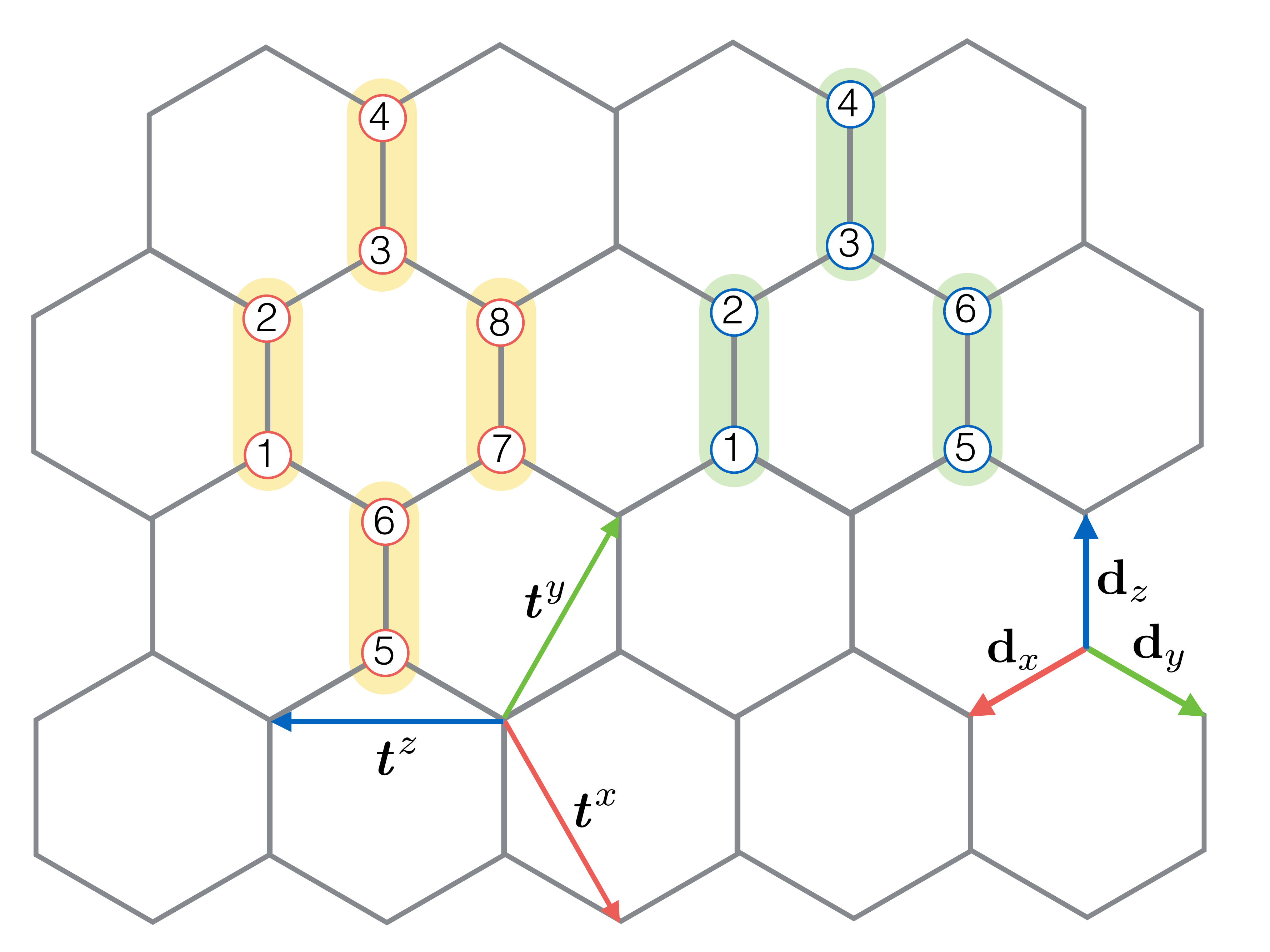}
\caption{(a) The honeycomb lattice with three types
of nearest neighbor bonds. Here  ${\bm t}^x = (\frac{1}{2}, \frac{-\sqrt{3}}{2})$, ${\bm t}^y = (\frac{1}{2}, \frac{\sqrt{3}}{2})$ are two primitive translations.
(b) Extended magnetic unit cells used in our variational calculation of the KH model. The quadrupoled unit cell (yellow shaded sites) corresponding to the general ordering composed of three wavevecotrs $\mathbf Q_1 = (-\pi, -\pi/\sqrt{3})$, $\mathbf Q_2 = (0, 2\pi/\sqrt{3})$, and $\mathbf Q_3 = (+\pi, -\pi/\sqrt{3})$. The tripled unit cell (green shaded sites), on the other hand, describes the $\sqrt{3} \times \sqrt{3}$ type ordering with a wavevector $\mathbf K = (4\pi/3, 0)$.
\label{fig:honeycomb} }
\end{figure}

As field is decreased, spins start to deviate from the $\hat{\mathbf n}$ direction. We next introduce a two-component vector $\bm \sigma_i = (\sigma^a_i, \sigma^b_i)$ and write the spin field as
\begin{eqnarray}
	\mathbf S_i = \sqrt{1 - |\bm\sigma_i|^2} \,\hat{\mathbf n} + \sigma^a_i \, \hat{\mathbf e}_a + \sigma^b_i \, \hat{\mathbf e}_b.
\end{eqnarray}
It is then easy to see that the individual spin component can be expressed as
\begin{eqnarray}
	S^\gamma_i = \frac{1}{\sqrt{3}} \,\sqrt{1 - |\bm\sigma_i|^2} + \sqrt{\frac{2}{3}}\, \bm\sigma_i \cdot {\bm t}^\gamma ,
\end{eqnarray}
 where ${\bm t}^x = (\frac{1}{2}, \frac{-\sqrt{3}}{2})$, ${\bm t}^y = (\frac{1}{2}, \frac{\sqrt{3}}{2})$, and ${\bm t}^z = (-1, 0)$ are the lattice vectors (see Fig.~\ref{fig:honeycomb}). Using this expression, we expand the spin interaction term $S^\gamma_i S^\gamma_j$ to second order in $\sigma$:
\begin{eqnarray}
	\label{eq:kitaev-term}
	& & S^\gamma_i \, S^\gamma_j = \frac{1}{3} \left(1 - \frac{|\bm\sigma_i|^2}{2} - \frac{|\bm\sigma_j|^2}{2} \right) \\ \nonumber
	& & \qquad\quad + \frac{\sqrt{2}}{3} \bm t^\gamma \cdot \left(\bm\sigma_i + \bm\sigma_j \right) + \frac{2}{3} (\bm\sigma_i \cdot {\bm t}^\gamma) (\bm\sigma_j \cdot {\bm t }^\gamma),
\end{eqnarray}
In particular, the isotropic Heisenberg exchange interaction $\mathbf S_i \cdot \mathbf S_j = \sum_{\gamma} S^\gamma_i S^\gamma_j$ becomes
\begin{eqnarray}
	\mathbf S_i \cdot \mathbf S_j = \left(1 - \frac{|\bm \sigma_i|^2}{2} - \frac{|\bm\sigma_j|^2}{2} \right)  + \bm\sigma_i \cdot \bm\sigma_j.
\end{eqnarray}

Substituting these expressions into the KH Hamiltonian, we obtain
\begin{eqnarray}
	\label{eq:quadratic-sigma}
	& &\mathcal{H} = E_0 + \frac{1}{2} (H - 3 J - 2 K) \sum_i |\bm\sigma_i |^2 \\ \nonumber
	& & %\quarticad
	 \,\, + J \sum_{\langle ij \rangle} \bm\sigma_i \cdot \bm\sigma_j 
	+ \frac{4K}{3} \sum_{\langle ij \rangle_\gamma} (\bm\sigma_i \cdot {\bm t}^\gamma) (\bm\sigma_j \cdot {\bm t}^\gamma).
\end{eqnarray}
where $E_0 = (3 J + 2 K) N - 2 H N$, and $N$ is the number of unit cells of the honeycomb lattice.
The terms linear in $\sigma$ in Eq.~(\ref{eq:kitaev-term}) cancel each other in the lattice sum. 
We note that the Hamiltonian~Eq.~(\ref{eq:quadratic-sigma}) can serve as a starting point for the quantum mechanical treatment of the magnon condensation. The spin ``deviations'' $\sigma^{a,b}_i$ are now quantum operators satisfying the commutation relations $[\sigma^a_i, \sigma^a_j] = [\sigma^b_i , \sigma^b_j ] = 0$, and $[\sigma^a_i , \sigma^b_j ] = i S \delta_{ij}$. In fact, the Holstein-Primarkoff boson operators are expressed as $a_i = (\sigma^a_i + i \sigma^b_i) / \sqrt{2 S}$. The magnon bandstructure is then obtained by diagonalizing the resultant magnon Hamiltonian using the Bogoliubov transformation. Magnetic instability occurs when one of the magnon bands touches zero as the field strength is decreased. 

Here we treat the spin deviations $\bm\sigma_i$ as classical variables and simply analyze the eigenmodes of the corresponding classical Hamiltonian. In particular, this classical instability analysis provides a direct comparison with the classical Monte Carlo simulations presented in the main text. To this end, we introduce Fourier transformation $\bm\sigma_i = \frac{1}{\sqrt{N}} \sum_{\mathbf k} \bm\sigma_s(\mathbf k) e^{i \mathbf k \cdot \mathbf r_i}$ to diagonalize the quadratic Hamiltonian Eq.~(\ref{eq:quadratic-sigma}). Here each site $i = (\mathbf r, s)$ is labeled by the Bravais lattice point $\mathbf r$ and the sublattice index $s = 1, 2$, $\mathbf r_i = \mathbf r + \mathbf d_s$ is the actual physical position of site-$i$, $\mathbf r = n_1 {\bm t}^x + n_2 {\bm t}^y$ are Bravais lattice points, and $\mathbf d_{1} = (0,0)$ and $\mathbf d_{2} = \mathbf d_{z} =(0,\frac{1}{\sqrt{3}})$ are basis vectors for the two sublattices.  The lattice geometry is shown in Fig.~\ref{fig:honeycomb}.

\begin{figure}[b]
\centering
\includegraphics[width=0.52\columnwidth]{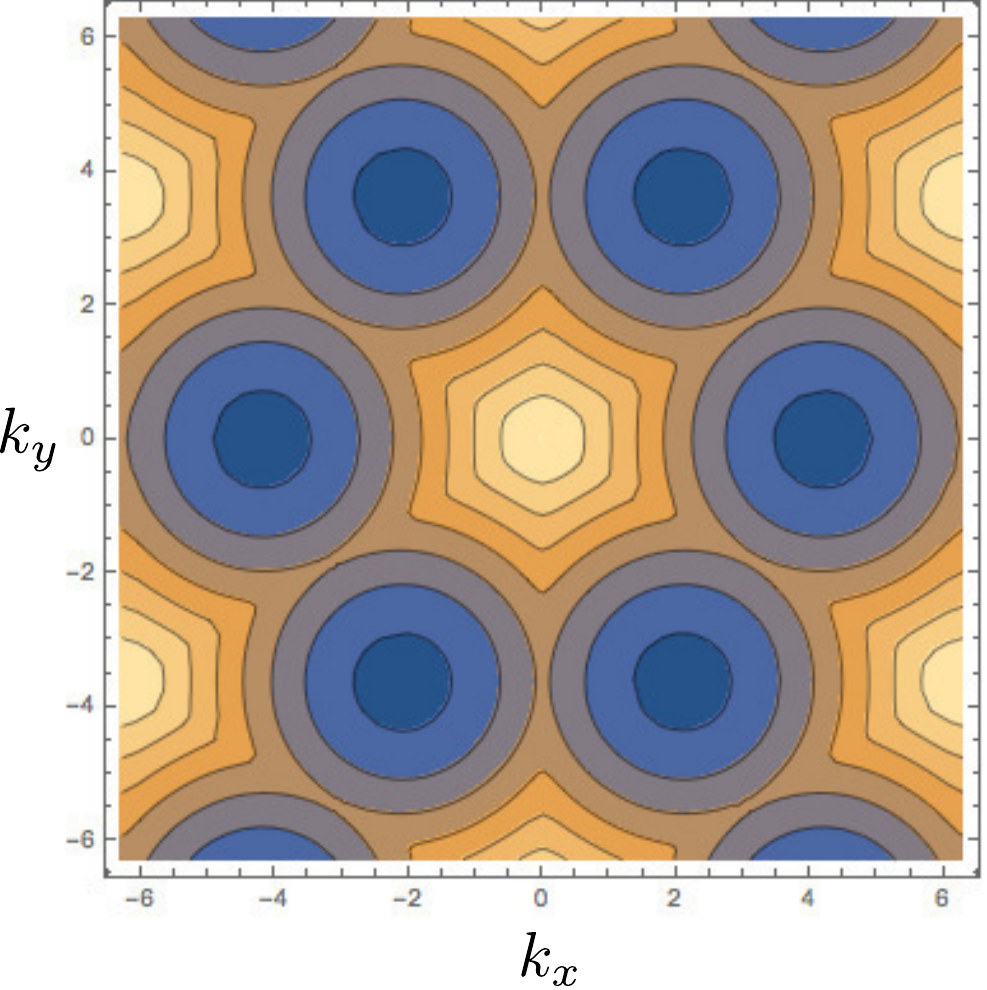}
\caption{Contour plot of the minimum eigenvalue of $\mathbb H(\mathbf k)$, showing minimum at the $K$ points $\mathbf Q_K = (\frac{4\pi}{3}, 0)$.
\label{fig:hk_min} }
\end{figure}

Substituting the Fourier expansion into Eq.~(\ref{eq:quadratic-sigma}), the spin Hamiltonian becomes
\begin{eqnarray}
	\mathcal{H} = E_0 + \sum_{\mathbf k} {\mathbb U}^*_{\mathbf k}\cdot  \mathbb{H}_{\mathbf k} \cdot {\mathbb U}_{\mathbf k}
\end{eqnarray}
where the 4-component vector ${\mathbb U}_{\mathbf k}= [\sigma^a_{1{\mathbf k}}, 
\sigma^b_{1{\mathbf k}} , \sigma^a_{2{\mathbf k}} , \sigma^b_{2{\mathbf k}} ]$. The interaction matrix $\mathbb H_{\mathbf k}$ has the following form:
\begin{eqnarray}
	\mathbb H_{\mathbf k} = \left[\begin{array}{cccc}
	\varepsilon_H & 0 & f_{\mathbf k} + g^{aa}_{\mathbf k} & g^{ab}_{\mathbf k} \\
	0 & \varepsilon_H & g^{ab}_{\mathbf k} & f_{\mathbf k} + g^{bb}_{\mathbf k} \\
	f_{-\mathbf k} + g^{aa}_{-\mathbf k} & g^{ab}_{-\mathbf k} & \varepsilon_H & 0 \\
	g^{ab}_{-\mathbf k} & f_{-\mathbf k} + g^{bb}_{-\mathbf k} & 0 & \varepsilon_H 
	\end{array} \right]. \,\,\,\,\,
\end{eqnarray}
The matrix elements are 
\begin{eqnarray}
	\varepsilon_H &=& \frac{1}{2}(H - 3J - 2K), \\
	f_{\mathbf k} & = &  = \frac{J}{2} \left( e^{i \mathbf k \cdot \mathbf d_x} +  e^{i \mathbf k \cdot \mathbf d_y} +  e^{i \mathbf k \cdot \mathbf d_z} \right), \\
	g^{aa}_{\mathbf k} & = & \frac{K}{6} \left( e^{i \mathbf k \cdot \mathbf d_x} +  e^{i \mathbf k \cdot \mathbf d_y} + 4 e^{i \mathbf k \cdot \mathbf d_z}\right), \\
	g^{bb}_{\mathbf k} & = & \frac{K}{2} \left( e^{i \mathbf k \cdot \mathbf d_x} +  e^{i \mathbf k \cdot \mathbf d_y} \right), \\
	g^{ab}_{\mathbf k} & = & - \frac{K}{2 \sqrt{3}} \left( e^{i \mathbf k \cdot \mathbf d_x} -  e^{i \mathbf k \cdot \mathbf d_y} \right).
\end{eqnarray}
Here the three vectors $\mathbf d_{x,y} = (\pm \frac{1}{2}, \frac{-1}{2\sqrt{3}})$, and $\mathbf d_z = (0, \frac{1}{\sqrt{3}})$ connect nearest-neighbors in honeycomb lattice.
As the field strength $H$ is reduced, the magnetic instability starts at the $\mathbf k^*$ points at which $\lambda_{\rm min}(\mathbf k^*)$ touches zero; here $\lambda_{\rm min}(\mathbf k)$ is the smallest eigenvalue of the matrix $\mathbb H(\mathbf k)$.  Figure.~\ref{fig:hk_min} shows the contour plot of $\lambda_{\rm min}(\mathbf k)$ in $k$-space. As can be seen, the function $\lambda_{\rm min}(\mathbf k)$ has minima at the $K$ points, indicating that the instability will take place at the corner of the Brillouin zone. The resultant  $\sqrt{3}\times \sqrt{3}$ magnetic ordering is consistent with our Monte Carlo simulation results at high field.

\begin{center}
{\bf II. Variational ground states}
\end{center}
%\label{sec:variational}

In this section we present a variational calculation for the classical ground states of the KH Hamiltonian. We consider magnetic structures with both a quadrupled unit cell and a tripled unit cell as our ansatz; see Fig.~\ref{fig:honeycomb}. In the former case, the 8-site spin structure includes the simple ferromagnetic and N\'eel orders with $\mathbf Q_0 = 0$, as well as the general zigzag and stripe orders characterized by wavevectors $\mathbf Q_1 = (-\pi, -\pi/\sqrt{3})$, $\mathbf Q_2 = (0, 2\pi/\sqrt{3})$, and $\mathbf Q_3 = (+\pi, -\pi/\sqrt{3})$. As discussed in the previous section, magnetic instability from the saturated state starts at the $\mathbf K = (4\pi /3, 0)$ points of the BZ. The corresponding eigen-mode belongs to the class of magnetic states with a tripled unit cell containing 6 inequivalent spins. In both cases, each spin in the extended unit cell is parameterized by two angles: $\mathbf S_i = S (\sin\beta_i \cos\alpha_i, \sin\beta_i \sin\alpha_i, \cos\beta_i)$. The total variational energy $E_{\rm var}(\{\alpha_i, \beta_i\})$, which is a function of these angle variables, is then minimized to obtain the variational ground states.

Next we discuss the characterization of the minimum-energy solution in the quadrupled unit cell. We first define vector order parameters that correspond to wavevector $\mathbf Q_0$ and the three $\mathbf Q_m$ ($m = 1, 2, 3$) at the $M$-points of the BZ. By labelling the 8 inequivalent sites according to Fig.~\ref{fig:honeycomb}, these vector order parameters are basically linear transformations of the eight spins $\{\mathbf S_i\}$:
\begin{eqnarray}
	\mathbf M &= \frac{1}{8}(\mathbf S_1 + \mathbf S_2 + \mathbf S_3 + \mathbf S_4 + \mathbf S_5 + \mathbf S_6 + \mathbf S_7 + \mathbf S_8), \,\,\,\, \nonumber \\
	\mathbf L &= \frac{1}{8}(\mathbf S_1 - \mathbf S_2 + \mathbf S_3 - \mathbf S_4 + \mathbf S_5 - \mathbf S_6 + \mathbf S_7 - \mathbf S_8), \,\,\,\,  \nonumber \\
	\mathbf N_1 & =  \frac{1}{8}(\mathbf S_1 - \mathbf S_2 - \mathbf S_3 + \mathbf S_4 + \mathbf S_5 - \mathbf S_6 - \mathbf S_7 + \mathbf S_8), \,\,\,\, \nonumber \\
	\mathbf N_2 & =  \frac{1}{8}(\mathbf S_1 + \mathbf S_2 - \mathbf S_3 - \mathbf S_4 - \mathbf S_5 - \mathbf S_6 + \mathbf S_7 + \mathbf S_8), \,\,\,\, \nonumber \\
	\mathbf N_3 & =  \frac{1}{8}(\mathbf S_1 - \mathbf S_2 + \mathbf S_3 - \mathbf S_4 - \mathbf S_5 + \mathbf S_6 - \mathbf S_7 + \mathbf S_8), \,\,\,\,   \\
	\mathbf R_1 & =  \frac{1}{8}(\mathbf S_1 + \mathbf S_2 - \mathbf S_3 - \mathbf S_4 + \mathbf S_5 + \mathbf S_6 - \mathbf S_7 - \mathbf S_8), \,\,\,\,  \nonumber\\
	\mathbf R_2 & =  \frac{1}{8}(\mathbf S_1 - \mathbf S_2 - \mathbf S_3 + \mathbf S_4 - \mathbf S_5 + \mathbf S_6 + \mathbf S_7 - \mathbf S_8), \,\,\,\,  \nonumber\\
	\mathbf R_3 & =  \frac{1}{8}(\mathbf S_1 + \mathbf S_2 + \mathbf S_3 + \mathbf S_4 - \mathbf S_5 - \mathbf S_6 - \mathbf S_7 - \mathbf S_8). \,\,\,\, \nonumber
\end{eqnarray}
Here the $\mathbf Q_0 = 0$ part includes $\mathbf M$, which is the simple ferromagnetic order, and $\mathbf L$ which describes the staggering of sublattice magnetization. The vectors $\mathbf N_m$ characterize the odd-parity zigzag order with wavevectors~$\mathbf Q_m$. And finally, the even-parity combinations corresponding to the stripe order are given by the three vector parameters~$\mathbf R_m$.
For spin Hamiltonians that preserve the SU(2) or O(3) spin rotational symmetry, or if the spin rotations are decoupled from the real-space symmetry operations, these vectors are the appropriate order parameters for the characterization of the magnetically ordered states. 

However, the presence of the anisotropic Kitaev term in the KH Hamiltonian explicitly breaks the spin rotational symmetry, and only generalized symmetry operations that involve discrete rotations in both spatial and spin spaces are preserved. For example, permutations of the three vector parameters $\mathbf N_m$ (by the $C_3$ rotations) must be accompanied by the corresponding rotation in spin space. Consequently, instead of the vector parameters listed above, the proper ordering parameters are given by the irreducible representations of the group of combined symmetry operations. For instance, as discussed in the main text, a multiple-$\mathbf Q$ zigzag order is characterized by a triplet of Ising parameters $\bm\phi = (\phi_1, \phi_2, \phi_3)$. Similarly, a multiple-$\mathbf Q$ stripe order is described by a triplet $\bm\eta = (\eta_1, \eta_2, \eta_3)$. In terms of these Ising order parameters, the corresponding vector parameters are $\mathbf N_m = \phi_m \, \hat{\mathbf e}_\gamma$ and $\mathbf R_m = \eta_m \, \hat{\mathbf e}_\gamma$. Here $m = 1, 2, 3$ corresponds to $\gamma = x$, $y$, $z$. 
Our direct numerical minimization finds that combined $C_3$ symmetry is preserved in the variational ground states in the parameter regime of our interest. As a result, for example, the symmetric zigzag order with $\phi_1 = \phi_2 = \phi_3$ is specified by only one scalar parameter.

\begin{figure}[t]
\centering
\includegraphics[width=0.82\columnwidth]{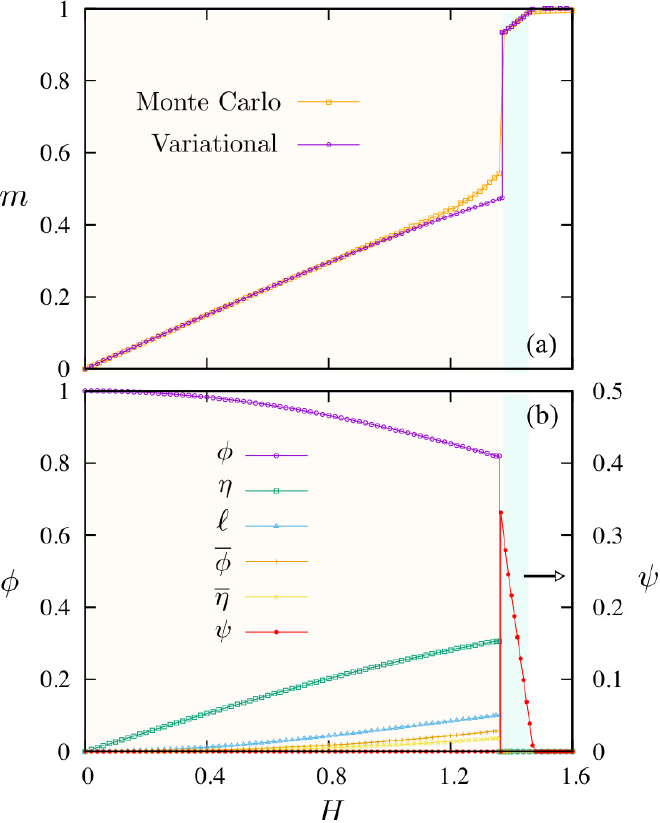}
\caption{
\label{fig:h_dep_var} 
Variational ground-state calculation of KH model at $\varphi = 0.7 \pi$ in the  magnetic field along the  $[111]$ direction : (a) Magnetization given by the ferromagnetic order parameter $m$ as a function of field strength. Also shown for comparison is the magnetization curve obtained from Monte Carlo simulations at a temperature $T = 0.005$. (b) The amplitude of the various order parameters defined in Eqs.~(\ref{eq:M-param})~and~(\ref{eq:K-param}) versus $H$. $T$ and $H$ are measured in units of $A$. }
\end{figure}

In the limit of $H \to 0$, the only nonzero order parameters are the three vectors $\mathbf N_m$  while all other vectors vanish. The magnetic field not only induces a finite magnetization $\mathbf M$, but also generates other small secondary order parameters due to the hard constraint of fixed spin length $|\mathbf S_i| = S$.  Through our direct numerical minimization, we find that the variational ground state of the KH model can be described by six scalar parameters $m$, $\ell$, $\phi$, $\bar\phi$, $\eta$, and $\bar\eta$:
\begin{eqnarray}
	\label{eq:M-param}
	\mathbf M &=& m \,(\hat{\mathbf e}_x + \hat{\mathbf e}_y + \hat{\mathbf e}_z) /\sqrt{3}, \nonumber\\ \nonumber
	\mathbf L &=& \ell\, (\hat{\mathbf e}_x + \hat{\mathbf e}_y + \hat{\mathbf e}_z) /\sqrt{3}, \\\nonumber
	\mathbf N_1 &=& \phi \, \hat{\mathbf e}_x /\sqrt{3} + \bar \phi \, (\hat{\mathbf e}_y + \hat{\mathbf e}_z) /\sqrt{6}, \\\nonumber
	\mathbf N_2 &=& \phi \, \hat{\mathbf e}_y /\sqrt{3} + \bar\phi\, (\hat{\mathbf e}_z + \hat{\mathbf e}_x) /\sqrt{6} , \\
	\mathbf N_3 &=& \phi \, \hat{\mathbf e}_z /\sqrt{3}  + \bar \phi \, (\hat{\mathbf e}_x + \hat{\mathbf e}_y) /\sqrt{6}  , \\\nonumber
	\mathbf R_1 &=& \eta \, (\hat{\mathbf e}_y + \hat{\mathbf e}_z)/\sqrt{6} + \bar \eta\, \hat{\mathbf e}_x /\sqrt{3}, \\\nonumber
	\mathbf R_2 &=& \eta\, (\hat{\mathbf e}_z + \hat{\mathbf e}_x) /\sqrt{6} + \bar \eta \, \hat{\mathbf e}_y /\sqrt{3} , \\\nonumber
	\mathbf R_3 &=& \eta \, (\hat{\mathbf e}_x + \hat{\mathbf e}_y)/\sqrt{6}  + \bar \eta \, \hat{\mathbf e}_z /\sqrt{3} , 
\end{eqnarray}
With these variational parameterization, the energy density of the 8-site spin structure is 
\begin{eqnarray}
	\varepsilon &=& -H m - \frac{1}{2} (3J + 2K) (\ell^2 - m^2) \nonumber \\
	& & + \frac{J}{2} \left( \phi^2 + \bar \phi^2 - \eta^2 - \bar\eta^2 \right) \\
	& & - K \left(\phi^2 - \bar \phi^2 + \eta^2 - \bar \eta^2 \right). \nonumber
\end{eqnarray}
The two exchange terms of the KH Hamiltonian are parameterized as $J = A \cos\varphi$, and $K = A \sin\varphi$. 
For a strong ferromagnetic Kitaev interaction ($K > 0$), as in the case of KH parameter $\varphi = 0.7 \pi$, the two dominant orderings are zigzag order characterized by $\phi$ and the stripe order characterized by $\eta$. The zigzag pattern is further favored by a antiferromagnetic Heisenberg term with $J < 0$, again as in the case of $\varphi = 0.7 \pi$. Indeed, as shown in Fig.~\ref{fig:h_dep_var},  a significant stripe order $\eta$ appears at high field in addition to the dominant zigzag order $\phi$. Finally, we note that the N\'eel order $\ell$ and $\bar \phi$, $\bar \eta$ are secondary parameters with small amplitude.

We next turn to the characterization of the magnetic structure with tripled unit cell. Other than the usual ferromagnetic $\mathbf M$ and N\'eel order $\mathbf L$, we are most interested in the order parameter corresponding to the $\sqrt{3} \times \sqrt{3}$ type pattern. This long-range order is characterized by a wavevector $\mathbf K = (4\pi/3, 0)$. For convenience, we define $\omega = \exp( i \,\mathbf K \cdot {\mathbf t}^x) = \exp(i \,2\pi/3)$. Using the labeling of the six inequivalent spins in Fig.~\ref{fig:honeycomb}, the appropriate vector order parameters are then given by 
\begin{eqnarray}
	\mathbf V_1 &= \frac{1}{3} \,(\mathbf S_1 + \omega \, \mathbf S_3 + \omega^2 \, \mathbf S_5), \nonumber \\
	\mathbf V_2 &= \frac{1}{3} \, (\mathbf S_2 + \omega\, \mathbf S_4 + \omega^2 \, \mathbf S_6). 
\end{eqnarray}
Here the subscript 1, 2 refers to the two sublattices of the honeycomb lattice. Consistent with the linear stability analysis discussed in the previous section, we find that the $\sqrt{3}\times \sqrt{3}$ structure indeed has a lower energy compared with the general 8-site ansatz in the high field regime. Moreover, our direct minimization shows that the $\sqrt{3}\times \sqrt{3}$ order can be characterized by a complex order parameter $\psi$ as follows:
\begin{eqnarray}
	\label{eq:K-param}
	\mathbf V_1 &=& +\psi \, (\hat{\mathbf e}_x + \omega \, \hat{\mathbf e}_y + \omega^2 \, \hat{\mathbf e}_z ) , \nonumber\\
	\mathbf V_2 & =& -\psi \, (\omega^2 \hat{\mathbf e}_x + \omega \, \hat{\mathbf e}_y + \hat{\mathbf e}_z) ,
\end{eqnarray}
where the phase of $\psi$ is field dependent. 
Fig.~\ref{fig:h_dep_var} summarizes our numerical calculation of the variational ground states. Other than the fully polarized state at high field, there are two nontrivial ordered states separated by a first-order phase transition at $H_c \sim 1.37 $. The low-field phase is the symmetric triple-$\mathbf Q$ order with a dominant zigzag order parameter~$\phi$. While the only nonzero order at $H \to 0$ is given by $\phi = 1$, all other order parameters are induced by the magnetic field and grow gradually with increasing $H$. Interestingly, a small N\'eel order is generated by the field. Moreover, the stripe order characterized by $\eta$ becomes quite significant in the intermediate field regime. For field strength above $H_c$, all order parameters related to three $\mathbf Q_\gamma$ wavevectors suddenly drop to zero. The high-field ground state corresponds to a finite $\psi$, indicating the $\sqrt{3}\times \sqrt{3}$ type long-range order.

We note  that the variational ground states are consistent with the Monte Carlo simulations for regimes where the ground state is the commensurate triple-$\mathbf Q$ zigzag (small $H$), and the $\sqrt{3}\times\sqrt{3}$ order (large $H$). The two methods give very consistent values for the $H_c$ of the first-order transition and the saturation field; see the comparison in Fig.~\ref{fig:h_dep_var}(a). However, since the variational calculation is restricted to commensurate unit cells, it cannot address the commensurate-incommensurate transitions and the novel incommensurate zigzag orders observed in Monte Carlo simulations. The variational approach, nonetheless, provides a guideline of the underlying energetics and serves as a useful double check for the large-scale simulations. 

The triple-$\mathbf Q$ zigzag order has an interesting canting pattern shown in the animation Canting.gif attached in the supplementary material.
At $H = 0$, the eight inequivalent spins point in the eight symmetry-related $\langle 111 \rangle$ directions. As $H$ is increased, the two spins pointing along $[111]$ and $[{\bar 1}{\bar 1}{\bar 1]}$, are completely unaffected by the field. The other six spins cant towards the direction of the field, with the canting angle increasing as a function of the field magnitude. 
%Among the six canting spins there are two distinct groups: spins starting with two positive signs at the start  $[{\bar 1}11]$ and  $[1{\bar 1}1]$. The latter group can be seen lagging behind the former in the canting pattern. 
At intermediate field, this canted triple-$\mathbf Q$ zigzag gives way to the incommensurate zigzag orders, phases III and IV discussed in the main text. As discussed above, the variational calculation based on 8-sublattice unit cell cannot describe the corresponding C-IC transitions. 
Finally, at high enough magnetic field it is no longer energetically favorable to keep one  spin in the direction opposite of the field and the results of the calculation revert back to single-$\mathbf Q$ commensurate zigzag phase with canted spins from our variational calculation. However, it should be noted that this high-field two-sublattice zigzag is only a metastable state. 
As shown in Fig.~\ref{fig:h_dep_var}, the six-sublattice $\sqrt{3}\times \sqrt{3}$ order is the ground state in the field regime immediately below the saturation field.

%However, from Monte Carlo simulations the system goes to triple-Q partial incommensurate zigzag phase (III) instead.  As we discussed above,  this phase can not be captured using the variational approach, and thus instead jumps to the energetically lowest commensurate phase, the single-$\mathbf Q$ commensurate zigzag phase (I).

\begin{center}
{\bf III. Nematic order}
\end{center}

In this section, we characterize the various zigzag phases using the nematic order parameter. The nematic phase of liquid crystals is marked by a preferred direction of the molecules. While ordered magnetic phases such as ferromagnetic or N\'eel order give rise to a nonzero nematic order parameter, an intriguing possibility is a phase which breaks the rotational symmetry while preserving the time-reversal symmetry. Such a spin nematic phase has been discussed in several quantum and frustrated magnetic systems. Here we are interested in the so-called uniaxial order parameter as a measure of the collinearity of spins. Specifically, we first compute the second-rank tensor order parameter:
\begin{eqnarray}
	Q_{\alpha\beta} = \frac{3}{2} \langle S^\alpha \, S^\beta \rangle - \frac{1}{2} \delta_{\alpha\beta},
\end{eqnarray}
where $S^\alpha$ is the $\alpha$ component of spin.
The uniaxial order parameter $\lambda_Q$ is then given by the largest eigenvalue of a $3\times 3$ matrix whose elements correspond to the above second-rank tensor. A full collinear spin configuration, e.g., a ferromagnetic or N\'eel order, is characterized by a maximum $\lambda_Q = 1$, while a completely disordered state has a vanishing uniaxial order parameter.

\begin{figure}[t]
\centering
\includegraphics[width=0.9\columnwidth]{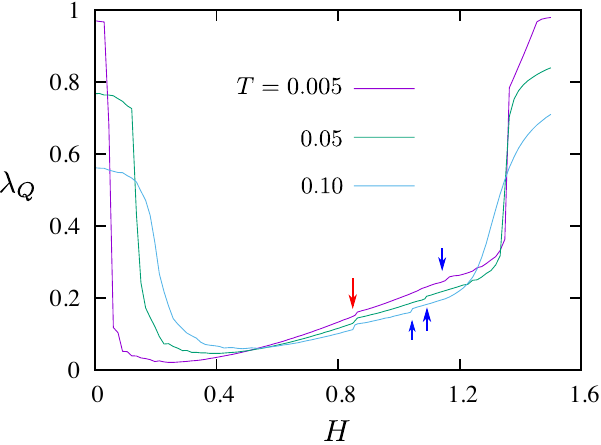}
\caption{\label{fig:nematic}
Field dependence of the uniaxial nematic order parameter $\lambda_Q$ at various temperatures. The arrows indicate the small jumps of $\lambda_Q$ at the commensurate-incommensurate phase transitions. $T$ and $H$ are measured in units of $A$.
}
\end{figure}

Fig.~\ref{fig:nematic} shows the field dependence of the uniaxial order parameter $\lambda_Q$ obtained from our Monte Carlo simulations for three different temperatures. As discussed in the main text, the low-temperature phase at small field is the collinear single-$\mathbf Q$ zigzag state. A rather large $\lambda_Q \approx 1$ in this regime is consistent with this conclusion. As $H$ is increased, the transition into the triple-$\mathbf Q$ zigzag phase is marked by a pronounced drop of the uniaxial order parameter as demonstrated in Fig.~\ref{fig:nematic}. In fact, the second-rank tensor $Q_{\alpha\beta}$ vanishes identically in a perfect triple-$\mathbf Q$ zigzag state. As the field strength is further increased, the tilting of spins toward the $[111]$ direction gradually increases the uniaxial parameter. Interestingly, $\lambda_Q$ exhibits small jumps at the two commensurate-incommensurate (C-IC) transitions, i.e. from zigzag phase II to III and from III to IV. Since thermal fluctuations tend to favor collinear spin configurations, the observed jumps of $\lambda_Q$ imply that the C-IC transitions might be partially driven by entropic selection. Finally, the transition from the zigzag phase IV to the $\sqrt{3}\times \sqrt{3}$ order at $H_c \sim 1.37$ is accompanied by a pronounced increase of the uniaxial order parameter.

\begin{figure}[t]
\centering
\includegraphics[width=0.99\columnwidth]{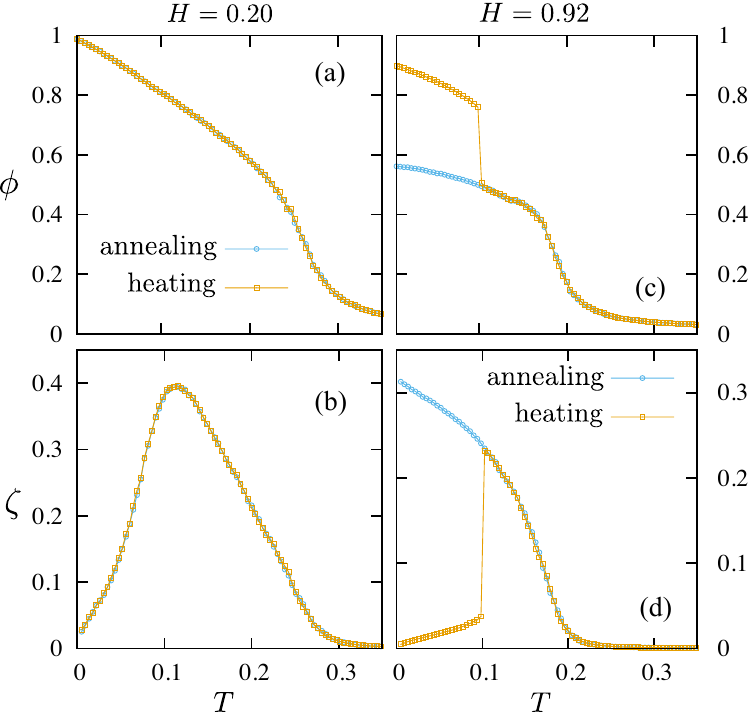}
\caption{\label{fig:T_dep}
Temperature dependence of order parameters $\phi = |\bm \phi|$ and $\zeta = |\bm\zeta|$ from annealing and heating simulations. Panels (a) and (b) are obtained with $H = 0.2$, while (c) and (d) are obtained with $H = 0.92$. 
}
\end{figure}

\bigskip

\begin{center}
{\bf IV. Temperature dependence and hysteresis}
\end{center}

Here we present the temperature dependence of the zigzag order parameter $\phi$ and $\zeta$. At small field, as shown in Fig.~\ref{fig:T_dep}(a) for $H = 0.2$, the zigzag order parameter $\phi$ increases monotonically as temperature is lowered. On the other hand, the amplitude of the doublet order parameter $\bm \zeta$ which characterizes the disparity of the three zigzag Ising parameters $\phi_m$ shows a non-monotonic temperature dependence; see Fig.~\ref{fig:T_dep}(b). As discussed in the main text, the doublet order parameter vanishes identically in a perfect triple-$\mathbf Q$ zigzag state, while $\zeta = |\bm\zeta|$ reaches its maximum value in a single-$\mathbf Q$ zigzag. The re-entrant behavior shown in Fig.~\ref{fig:T_dep}(b) thus corresponds to an intermediate single-$\mathbf Q$ zigzag phase that is stabilized by thermal fluctuations at finite temperatures. The absence of hysteresis from the annealing and heating simulations points to a continuous transition between the single and triple $\mathbf Q$ zigzag phases.

At high field $H = 0.92$, annealing simulation from a disordered state shows a monotonic growth for both order parameters $\phi$ and $\zeta$ with decreasing temperature; see Fig.~\ref{fig:T_dep}(c) and (d). From the $H$-$T$ phase diagram shown in the main text, there are two low-$T$ zigzag phases at this field value: the single-$\mathbf Q$ commensurate phase I and the partially incommensurate phase III at lowest temperatures. Since the $C_3$ symmetry is broken in both phases, the $\zeta$ order parameter describing the disparity of the three zigzag chains is nonzero throughout the low-$T$ ordered regime.  Interestingly, our simulations also find that the incommensurate zigzag phase III coexists with the commensurate triple-$\mathbf Q$ zigzag II state over a wide range of temperatures, as demonstrated by the pronounced hysteresis loop from the annealing and heating simulations shown in Fig.~\ref{fig:T_dep}(c) and (d). In the heating simulations, the spins are initialized to the commensurate triple-$\mathbf Q$ zigzag state obtained from the variational minimization discussed above. At zero temperature, this triple-$\mathbf Q$ phase with three coexisting zigzag Ising order parameters $\phi_1 = \phi_2 = \phi_3$ is characterized by a vanishing $\zeta$. As $T$ increases, we find that the triple-$\mathbf Q$ state is a very robust local minimum and remains stable until $T \sim 0.1$, above which the system decays spontaneously into the partially incommensurate zigzag phase III as indicated by a sudden increase of the $\zeta$ order parameter.

\end{document}